# ·A novel TOF-PET MRI detector for diagnosis and follow up of the prostate cancer

F. Garibaldi[1,2,*], S. Beging[9], R. Canese[3], G. Carpinelli[3], N. Clinthorne[4], S. Colilli[1,2], L. Cosentino[5], P. Finocchiaro[5], F. Giuliani[1,2], M. Gricia[1,2], M. Lucentini[1,2], S. Majewski[6], E. Monno[7], P. Musico[8], F. Santavenere[1,2], J.Tödter[9], H. Wegener[9], K. Ziemons[9]

1. INFN Roma Piazzale Aldo Moro 1 – Rome - Italy
2. ISS - National Center of Innovative Technologies for the Public Health - Viale Regina Elena 299 – 00161 – Rome – Italy
3. ISS - Cell Biology and Neurosciences Department-Rome - Viale Regina Elena 299 – 00161 – Rome – Italy
4. Div. Nuclear Medicine / Dept. Radiology, University of Michigan, Ann Arbor, Michigan, USA 48109-5610
5. INFN LNS – Vis S. Sofia 62 – 95125 Catania
6. University of Virginia, Charlottesville, Virginia, USA
7. ENEA- Centro ricerche di Brindisi – S.S. 7 Appia km 706,00 - 72100 Brindisi
8. INFN Genova – Via Dodecaneso 33 - 16146 Genova - Italy
9. FH Aachen University of Applied Sciences Heinrich-Mußmann-Strasse 1 52428 Jülich | Germany

**Abstract**

Prostate cancer is the most common disease in men and the second leading cause of death from cancer. Generic large imaging instruments used in cancer diagnosis have sensitivity, spatial resolution, and contrast inadequate for the task of imaging details of a small organ such as the prostate. In addition, multimodality imaging can play a significant role merging anatomical and functional details coming from simultaneous PET and MRI. Indeed, multi-parametric PET/MRI was demonstrated to improve diagnosis, but it suffers from too many false positives. In order to address the above limits of the current techniques, we have proposed, built and tested, thanks to the TOPEM project funded by Italian National Institute of Nuclear Phisics a prototype of an endorectal PET-TOF/MRI probe. In the applied magnification PET geometry, performance is dominated by a high-resolution detector placed closer to the source. The expected spatial resolution in the selected geometry is about 1.5 mm FWHM and efficiency a factor of 2 with respect to what obtained with the conventional PET scanner. In our experimental studies, we have obtained timing resolution of ~ 320 ps FWHM and at the same time Depth of Interaction (DOI) resolution of under 1 mm. Tests also showed that mutual adverse PET-MR effects are minimal. In addition, the matching endorectal RF coil was designed, built and tested. In the next planned studies, we expect that benefiting from the further progress in scintillator crystal surface treatment, in SiPM technology and associated electronics would allow us to significantly improve TOF resolution

1. Introduction

Prostate cancer is the most common disease in men and the second leading cause of cancer death [1]. Early diagnosis defining the exact tumor location(s), spread, structure (heterogeneity), and margins, would make targeted therapies more accurate, increasing the overall cure rates. Dedicated multimodality prostate imagers are expected to provide much better diagnostic power when compared to generic large imaging instruments by offering higher sensitivity, higher spatial resolution, and better contrast. For these reasons a prototype of a TOF-PET-MRI prostate probe was

---

designed and has been developed and tested, showing very promising initial results.

## 2. Present status of the management of the disease and applied diagnostic methods

Early diagnosis of prostate cancer (PCa) is difficult with the current diagnostic imaging devices. Current treatments treat the tumor as a whole instead of offering a patient-tailored approach at an early localized stage [1]. Early diagnosis defining the exact tumor location, spread, and margins, could make targeted therapies possible. This is expected to increase the overall survival while minimizing treatment side effects such as incontinence, impotence, etc. Imaging modalities represent important tools in PCa diagnostic work-up, both in the pre-treatment and in post-treatment phases. Most recent developments in imaging technologies (MRI and PET) can lead to significant improvements in both lesion detection and staging, and the image fusion with integrated anatomic and metabolic information can offer a potential advantage in comparison to the currently available diagnostic tools in PCa. Transrectal Ultrasound (TRUS) still remains the most widely applied tool to guide prostate biopsies [2]. However, its usefulness in local staging is limited because extracapsular extension (ECE) and seminal vesicle invasion (SVI) are difficult to visualize on ultrasound unless there is a large extension of the tumor. Bone scanning and CT are used for staging a more advanced disease and are useful only in patients at relatively high risk. More accurate staging would facilitate early treatment decisions and lead to a better outcome for patients. So new combinations of anatomic, functional, and molecular imaging approaches are needed. MRI has been largely investigated in evaluating intraprostatic disease, staging and restaging PCa; multiparametric MRI [mpMRI], including techniques such as diffusion-weighted imaging, dynamic contrast enhancement and magnetic resonance spectroscopy, improved diagnostic accuracy of MRI; nowadays, mpMRI is particularly good at accurately detecting anterior tumours that are usually missed by systematic biopsy and may be useful in selected patients with intermediate- to high-risk cancers but the accuracy is lower if performed within 2 months after biopsy. However, MR suffers from somewhat lower specificity [2,3,4]], and would be enhanced when combined with the molecular modalities such as high-resolution PET in simultaneously performed scans, as is proposed in this project. MRI preferentially detects higher Gleason grade cancers potentially making it a great screening tool for clinically relevant cancers—most Gleason 6 cancers can be considered indolent and will not affect patient outcome and therefore need no treatment [3]. Unfortunately, smaller poorly differentiated lesions may still be missed by mpMRI, particularly in the presence of multiple prostatic lesions [3]. Since the vast majority of men with prostate cancer do harbor multiple cancerous lesions in their prostates, and focal therapy of these lesions is an appealing means of reducing treatment toxicity [4], accurate detection of smaller higher-grade lesions is crucial for a proper selection of patients for focal treatment options. Metabolic and functional changes usually precede anatomical changes; functional imaging with Positron Emission Tomography (PET), improved with CT for anatomic information and attenuation correction (PET/Computed Tomography, PET/CT), provides additional details compared to conventional imaging [5]. However, a major disadvantage of prostate imaging with PET/CT is the limited soft tissue contrast provided by the CT scan, thereby impairing lesion targeting [6]. Furthermore, studies correlating MRI and surgical pathology of the specimens demonstrated a sensitivity of only 59% in identifying prostatic cancer [7] so this result is expected to be improved by PET in locus, adding information about the functional and metabolic changes of the gland cells.
For these reasons, we believe that a dedicated multi-parametric, high resolution high efficiency device is needed, such as the TOF-PET-MRI probe with biopsy capability proposed here.

## 3. The project TOPEM

We have developed an endorectal Time of Flight (TOF) PET - MRI probe. Exploiting the TOF capability provides an increase in the Signal to Noise Ratio (SNR) and in Noise Equivalent

Counting Rate (NECR) and also permits suppression of the bladder background [8,9].

Conventional external ring PET geometries suffer from two limitations for prostate imaging. First, since both annihilation photons must be detected in PET, attenuation of photons emitted from the centrally-located prostate in the surrounding body tissue is severe. Second, spatial resolution in PET images reconstructed from external ring scanners is relatively modest (~ 6 mm) and suboptimal to detect small few mm lesions. An external "ring" type PET scanner optimized for prostate imaging has been constructed [10,11]], using standard resolution block detectors directed toward the prostate, to minimize uncertainty due to unmeasured depth-of-interaction and with the ring placed close to the patient to maximize detection efficiency for a given detector size. Reconstructed resolution in phantom studies was reported at 4mm FWHM. While this is an incremental improvement over most commercially available instruments, photon attenuation is still an issue and image distortions outside the prostate, due to the scanner design, likely render detection of metastatic lesions in lymph nodes more difficult than with conventional scanners. For these reasons endorectal PET probes were discussed at the 2005 workshop on technology for prostate imaging held in Rome [12] where two relevant designs for optimized PET prostate imaging were presented. Both made use of an endorectally placed high resolution miniature PET detector working in coincidence with external standard resolution detectors. This asymmetric configuration provides higher resolution through the magnification geometry.

The resolution magnification concept has been well described in the literature. Magnifying PET geometries—where a detector having high spatial resolution located close to a region of interest works in coincidence with a conventional PET detector having more modest resolution—have been investigated in a number of studies spanning the past decade. Clinthorne and Park proposed instruments for small animal and patient imaging based on high-resolution detectors used in conjunction with standard PET detectors [13-17]. Tai and co-workers have referred to the concept as "virtual pinhole PET" and have developed several demonstration instruments [18-20]. Huh, and Clinthorne, have evaluated the concept of an endorectal insert for high resolution prostate imaging instrument [13]. Finally, in the most relevant experimental study, Stolin, and colleagues performed measurements on a Tandem Positron Emission Tomography System, composed of the panel and a probe and showing the advantages of such a detector system [21]. Other authors have considered the importance of such a concept [22] and/or performed simulations and measurements confirming its validity

The internal probe can be used in coincidence with external dedicated detectors and/or a standard PET ring. In order to be very close to the radiation source located in prostate, one of the detectors has to be placed in the endorectal probe. As discussed above, the performance of such an asymmetric structure for radiotracer activity in the prostate is dominated by the endorectal PET detector with improvements in both spatial resolution and efficiency [20,26].

A Geant 4 code has been written in order to perform simulations to optimize the detector layout. Fig. 1 shows the Zubal phantom [27]. used for the simulation.

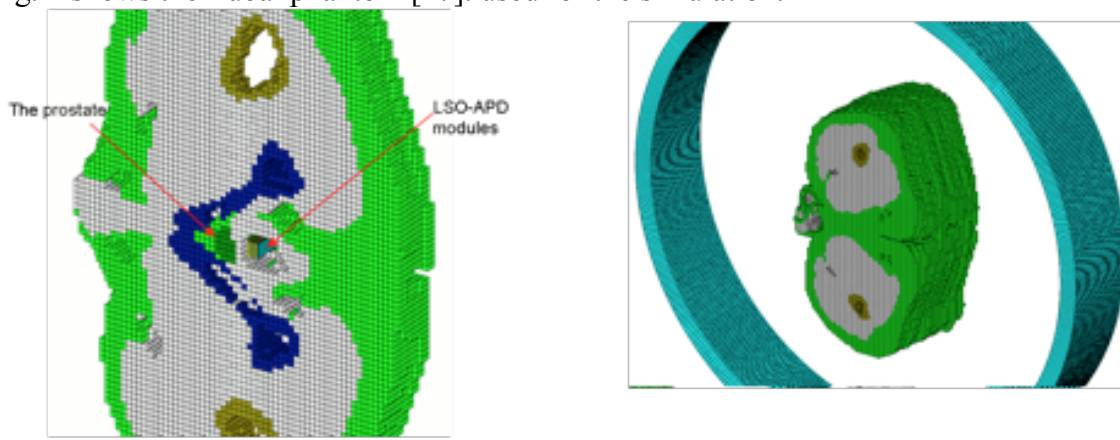

Fig. 1 Slice of the pelvic region containing the PET probe and a conventional PET scanner

The simulations (Fig 2) show that in the case of a standard PET scanner used in coincidence, spatial resolution of ~ 1.5 mm FWHM (for an internal probe resolution of 1 mm) can be obtained with such a system for source distances of 10–20 mm from the probe and that an increase in efficiency (and further enhanced by increasing the probe size) is also obtained.

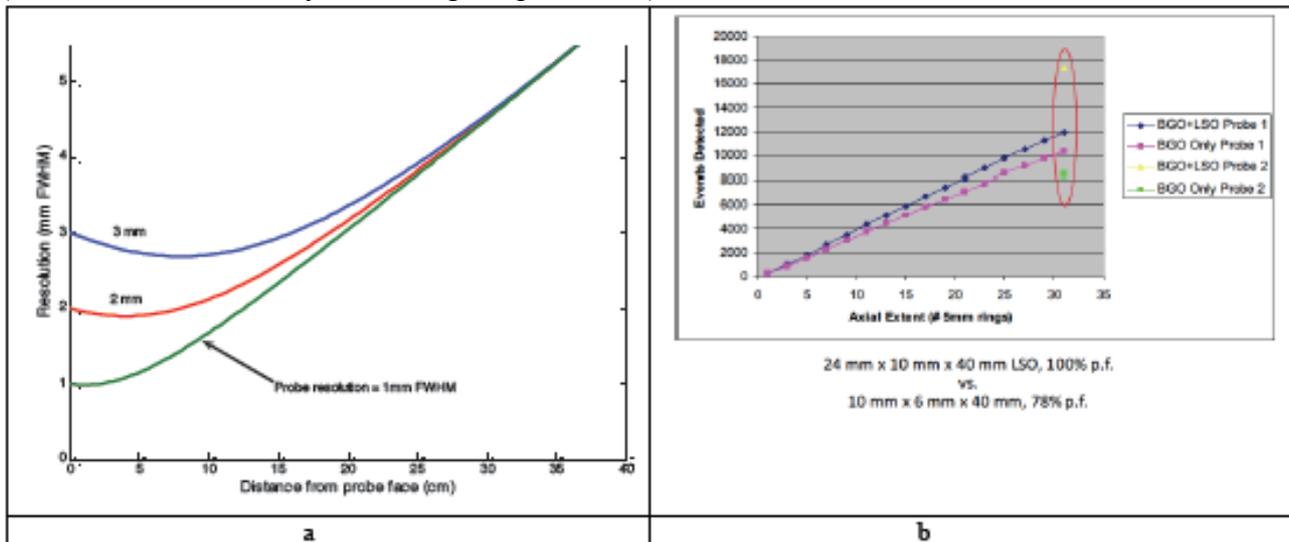

Fig. 2 Spatial resolution (a) and Efficiency (b) of the PET + probe system for different probe dimensions

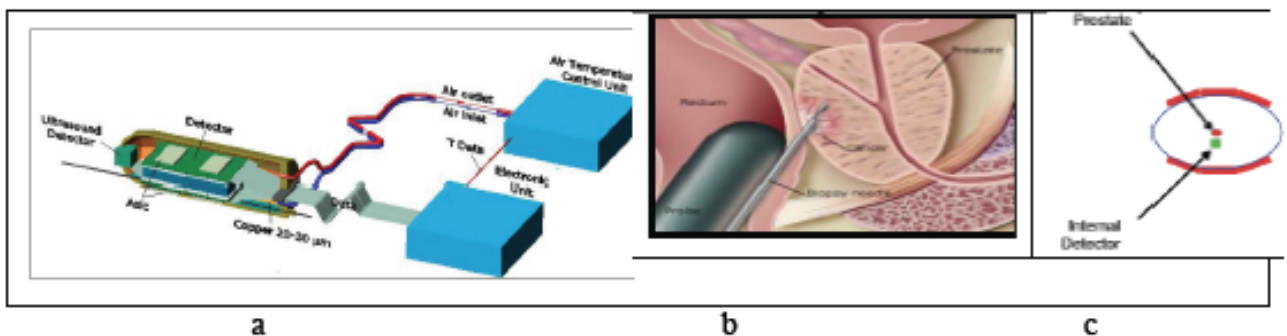

Fig. 3. a. Sketch of the probe with the electronics and the control system. b. The probe with the biopsy system. c. Layout of the diagnostic system: the probe in coincidence with external panels

Based on these prior simulation results, a design of the probe was prepared and is shown in Fig. 3 a. Fig. 3b shows the possible use of probe in biopsy.

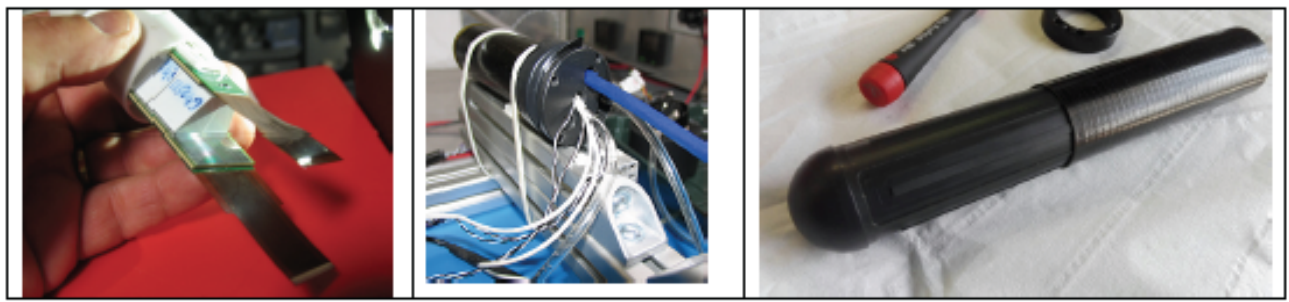

Fig.4 a. The probe with tapered pixelated LYSO scintillator sanwitched between SiPM photosensors. b. and c. the carbon fiber cage with the temperature stabilization system

Our choice was to design and build a system consisting of a probe + dedicated high resolution/high efficiency PET panels (Fig. 3c). In order to obtain artefact free images, 4 PET panels (200 x 200 x 20 mm3) should be built. Due to funding limitations, only a very small panel detector (50x50x20 mm$^3$) was built for the prototype system.

## 4. The probe

Based on the prior art and the above discussion, we have designed and built a probe prototype made of a pixelated LYSO scintillator array of 25 (width) x 13 (thickness) x 50 (length) mm$^3$ outside size with tapered layout (3 x 3 mm$^2$ pitch at the top and 1 x 2 mm$^2$ pitch at the bottom) to improve the efficiency while keeping good spatial resolution [28].

For compactness and MRI compatibility, Silicon Photomultipliers (SiPM) Hamamatsu (8 x8 arrays, 3 x 3 mm2 (top) and 1 x 1 mm2 (bottom)) were used as photosensors. Their intrinsic time jitter is almost negligible so the expected time resolution is a direct function of the square root of the number of produced photoelectrons.

In order to obtain good Depth Of Interaction (DOI) resolution, and to optimize light collection (important parameter affecting the TOF resolution), the scintillator pixels were coupled one to one to two sheets of SiPMs of the same dimension (see Fig.4).

To minimize the effect of the eddy currents due to the MRI scanner, copper shielding (25 μm in thickness) covers the interior part of the carbon fiber cylinder containing the probe,

Good timing resolution requires low and stable temperature for the electronics to minimize the noise and enable triggering on the first few photoelectrons. For this reason, a cooling system, with additional feedback on SiPM power supply is needed. The temperature of the detector and electronics were maintained at ~ 15 $^o$C while keeping the "external" part of the probe container, in contact with the rectum of the patient at ~ 36 $^o$C.

The carbon fiber cylinder has a hollow structure (Fig 4 c), allowing water circulation at a desired temperature and pressure to stabilize both internal (detector and electronics) and external temperature (see Fig.5).

A double-sided readout was implemented to have a good Depth of Interaction (DOI) resolution. Good timing resolution and good DOI resolution require different surface treatment and reflecting material. As discussed in the next paragraph, an optimal compromise has been found and further improvements are still possible

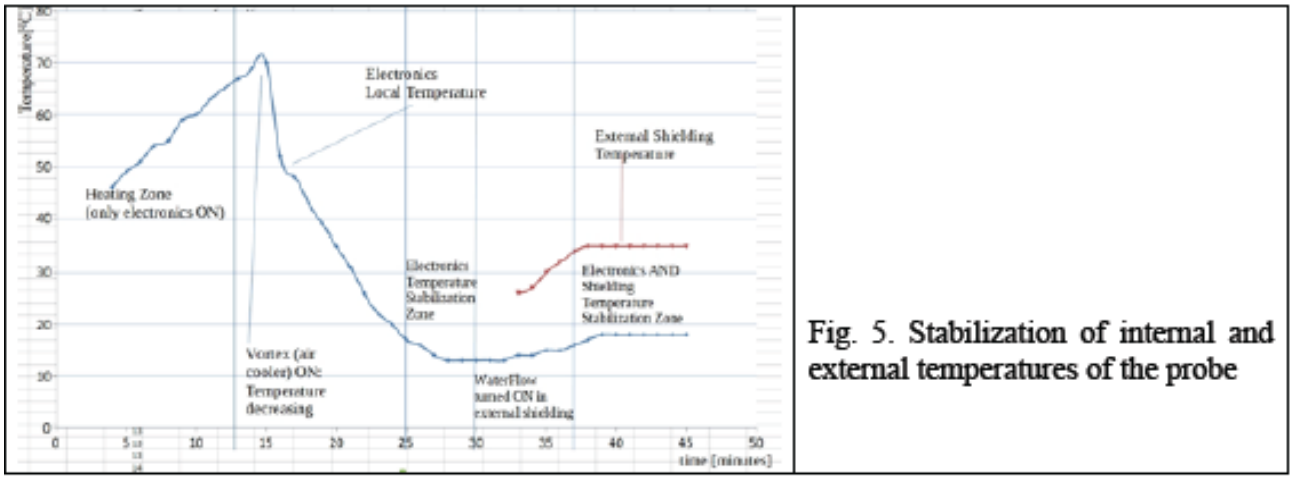

Fig. 5. Stabilization of internal and external temperatures of the probe

### 4.1 Time of Flight (TOF) and Depth of Interaction (DOI)

In our detector system, the distance of the probe to the panels is small with respect to a standard PET. For this reason, there could be problems of parallax errors (Fig. 6 a). The errors can be mitigated by measuring the position of interaction of the photons impinging on the scintillator crystals. This was achieved by measuring the charge asymmetry between the two photosensors installed on each crystal (Fig. 6 b).

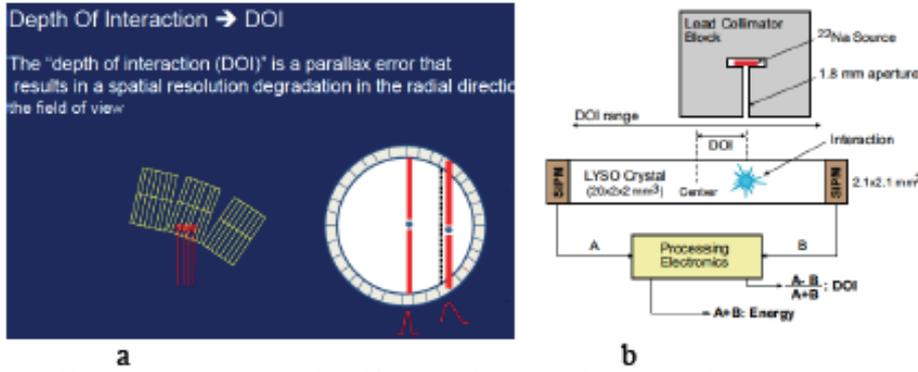

Fig 6. a. Parallax errors and Depth of interaction. b. The method used to measure DOI

Good time of flight resolution has several advantages. The major advantage of improved coincidence timing is the reduced random event rate [8]. The practical effect can be estimated using the noise equivalent count rate (NECR), a common figure of merit for comparing tomograph performances. The NECR value is given by a formula: NECR = $T^2/ (T+S+2R)$ where NECR is the noise equivalent count rate, T is the true coincidence event rate, S is the scattered event rate, and R is random event rate. Fig. 7 shows the improvement of SNR and NECR as a function of the timing resolution for different sizes of the object to be imaged [8]. Other advantages are the reduced scanning time and the possibility of getting rid of the artefacts from angular undersampling [19]. Timing resolution depends on several parameters like scintillator properties (light output, timing), path length spread, transit time in the photodetector, TDC time spread, discriminator time spread, system noise contributions (power supply, clocks, cables). All of them have to be optimized.

We performed measurements of the timing resolution that can be obtained by our device [30]. Measurements were performed with detector prototypes on the bench using standard nuclear physics laboratory electronics.

Both DOI and TOF optimization are needed. Unfortunately, this is difficult because of conflicting requirements (attenuation versus DOI thus losing photons on the rough scintillator surfaces; minimizing the light path thus maximizing the number of photons for TOF).
Therefore, the improvement of the DOI resolution implies that the time resolution worsens, as a consequence of the reduced number of detected photons. As part of our research, we performed measurements with a LYSO finger scintillator 1.5 x 1.5 x10 mm$^3$ with polished surfaces, wrapped with Lumirror [31] to have a slight loss of light propagating across the finger. In such a configuration the light loss due to the reflections is moderate, allowing the measurement of DOI with resolution around 1mm without significantly impacting the time resolution.
Two Hamamatsu SiPM photodetectors, with 3 x 3 mm$^2$ active surface and with 50μm microcell size, have been coupled to a LYSO scintillator by means of optical grease and a suitable mechanical fixture

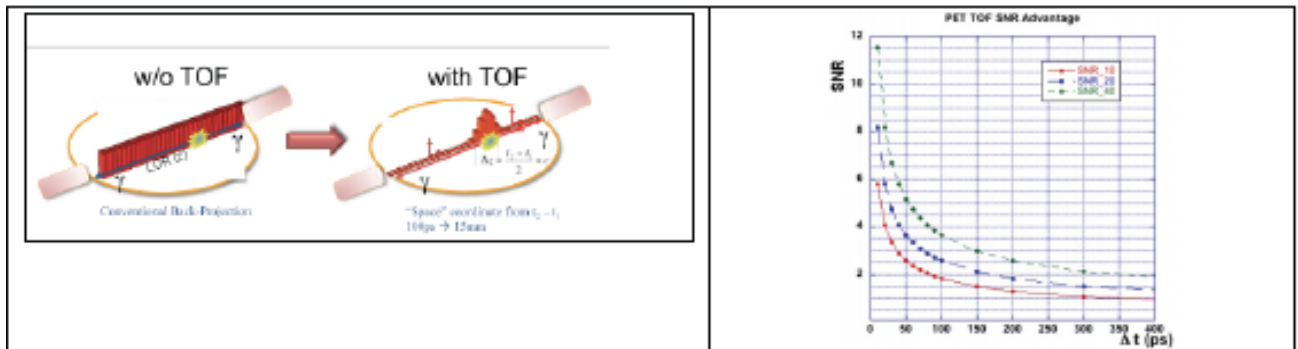

Fig. 7 Time of Flight (TOF) concept and advantages in terms of SNR for different object size (10, 20, 40 cm in diameter) and timing resolution (plot at right).

A lead collimator with hole diameter Φ = 1mm was used to irradiate the finger scintillator in five different points along its length with $^{137}$Cs and $^{22}$Na radioactive sources. In any irradiating position, the charge pulse by each SiPM and the difference between the arriving times on the finger sides have been acquired. In addition, a second detector made by having a standard PMT coupled to a small LYSO crystal (3x3x3 mm$^3$) has been used, operating in coincidence with the LYSO finger, by placing the point $^{22}$Na source in between, thus allowing the measurement of timing resolution. In Fig. 8, a sketch of the experimental apparatus is shown. Both SiPMs have been biased at 72.9V, the value allowing the best timing resolution, and the threshold of the leading edge discriminator has been set equivalent to 1.5 photon detected. The $^{22}$Na energy spectrum has been acquired by irradiating the scintillator without any collimator. The DOI was calculated event by event, by combining the charge read out by the SiPMs, DOI = **k** x charge1/(charge1+charge2) with **k** a calibration constant, and selecting only the photopeak events.

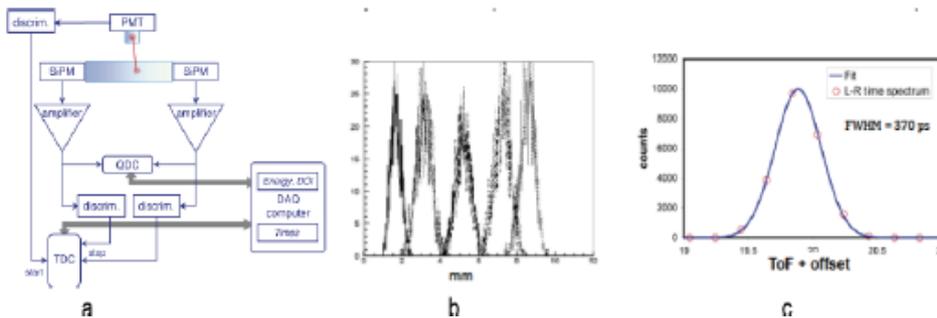

Fig 8 a. Sketch of the experimental apparatus. b. DOI distribution. c. TOF resolution

In Fig. 8 b, the DOI distributions corresponding to the five nominal positions (1.5, 3, 5, 7, 8.5 mm) acquired with the $^{22}$Na collimated source are shown; the FWHM of each peak is just below 1 mm. With the coincidence detector, the time of flight resolution of the gamma rays emitted in coincidence is 370 ps FWHM, (Fig. 8 c). Such a value, that is mainly affected by the worse timing resolution of the PMT detector in comparison with the LYSO finger, can be decreased down to 320 ps if the PMT is replaced by another double-sided readout LYSO finger. Considering that a 2-3 mm DOI resolution could still be acceptable for a required reconstructed spatial resolution of 1.5–2 mm, one could replace the Lumirror with a slightly better reflecting wrapping material, thus slightly increasing the photon statistics with a corresponding improvement in timing resolution.

We have not measured the spatial resolution and the efficiency of detector system (probe + external detector yet but simulations and measurements performed by other groups (two by authors of this paper) confirm the results of our simulation. Both spatial resolution and efficiency are significantly improved by our detector system with respect to a standard PET (or PET-MRI) scanner .

### 3.5 The readout electronics

Optimal electronics should detect and measure coincidences with a precision of 300 ps or less, and be small enough to be directly connected to the probe detector (see Fig. 4a).

The very compact electronics used in our prototype is based on a high performance ASIC (TOFPET ASIC [32]) featuring 64 channels, a TDC with 25 ps r.m.s. intrinsic time resolution, the Time Over Threshold technique, charge measurement, 300 pC dynamic range

Coincidence logic has been implemented, allowing interfacing with an external detector, trigger , and data acquisition electronics for implementing the TOF-PET detector system (probe + external PET detector(s)). The data acquisition software is based on Linux PC standard, interfaced to the controller board using FastEthernet connection. The controller uses a Field Programmable Gate Array (FPGA) to implement all needed logic functions, and can handle two SiPM arrays coupled to the pixelated scintillators of the probe for DOI measurement and the external detector(s).

Using the TOFPET ASIC a TOF resolution of ~ 240 ps FWHM has been obtained by another group [33]. Other results of measurements of TOF resolution by different groups show that it can be improved down to ~ 120 ps. [34].

## 5. PET – MR compatibility

We performed measurements to test the PET vs MRI compatibility. In order to quantify signal losses due to the introduction of the PET device into the MRI coil, a phantom was built. The phantom consisted in twelve NMR tubes filed with saline solution (in order to increase the load of the coil) positioned around a cylinder of 4 cm diameter in which the PET device could be positioned. In this way we were able to measure the signal present at different positions within the coil and around the device.

MRI experiments were performed on the phantom using a Varian INOVA MRI/MRS system (Varian, Palo Alto, USA) operating on small animals at 4.7 T (horizontal magnet, bore size 18.3 cm), equipped with actively shielded gradient coils (maximum gradient strength 200 mT/m; rise time <150 μs), with a transmitter and receiver volume RF coil (6 cm diameter, RAPID Biomedical, Rimpar, Germany). Contiguous multislice spin-echo images (TR/TE = 6000/20, matrix 128 x 128, Field of view = 60 x 60 mm, 2 averages) were collected in the axial plane as shown in Fig. 9a (so the tubes appear as white or gray circles).

MRI was acquired in different experimental conditions:
A) phantom with the carbon cage inside it,
B) phantom with the PET device switched off inside it,
C) phantom with the PET device switched on inside it.

The homogeneity of static magnetic fields was not optimal because of the large dimensions of the insert device compared to the inner volume of the coil. This is the reason for the different observed signal intensity among the tubes. The signal to noise ratio was calculated in all the experimental conditions. Signal intensities were taken from all the tubes (12) in all the slices (9) which are equally spaced of about 3.5 mm and two voxels for noise in each slice.

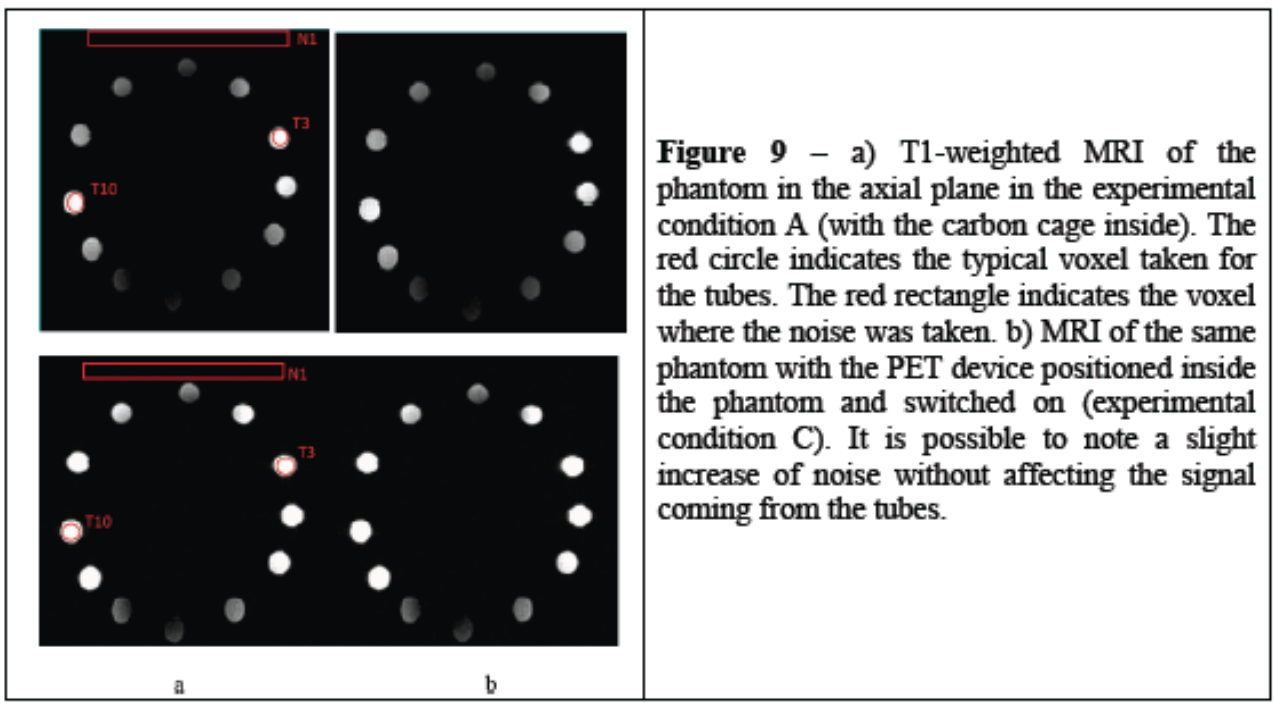

**Figure 9** – a) T1-weighted MRI of the phantom in the axial plane in the experimental condition A (with the carbon cage inside). The red circle indicates the typical voxel taken for the tubes. The red rectangle indicates the voxel where the noise was taken. b) MRI of the same phantom with the PET device positioned inside the phantom and switched on (experimental condition C). It is possible to note a slight increase of noise without affecting the signal coming from the tubes.

Signal to noise values for three representative tubes (tube 3, 9 and 12 which are positioned at different sites within the coil and have different signal intensity due to magnetic static field inhomogeneity) and taken from different voxels placed along the tubed (slice1, 5, and 9) are summarized in Table 1 and Table 2.

**Table 1** – Signal intensities coming from three different voxels (two from the tubes T3 and T10; a voxel for the noise N1), positioned in different slices, i.e. the center of the magnet (slice 5) and 1.6 mm before (slice 1) and after (slice 9) within the phantom in the A, B and C experimental conditions.

|   |   | T3 (a.u.) | T10 (a.u.) | N1 (a.u.) | T3/N1 | T10/N1 |
|---|---|---|---|---|---|---|
| A | slice 1 | 57.62 | 38.68 | 1.14 | 50.54 | 33.93 |
|   | slice5 | 61.46 | 39.75 | 1.21 | 50.79 | 32.85 |
|   | slice9 | 58.33 | 39.38 | 1.15 | 50.72 | 34.24 |
| B | slice 1 | 56.6 | 41.2 | 1.159 | 48.84 | 35.55 |
|   | slice5 | 55.73 | 41.6 | 1.191 | 46.79 | 34.93 |
|   | slice9 | 54.21 | 43.3 | 1.186 | 45.71 | 36.51 |
| C | slice 1 | 58 | 42.26 | 1.5 | 38.67 | 28.17 |
|   | slice5 | 57.78 | 43.07 | 1.505 | 38.39 | 28.62 |
|   | slice9 | 56.4 | 44.9 | 1.52 | 37.11 | 29.54 |

a.u. = arbitrary units.

**Table 2** – Percentage of signal losses due to the different experimental conditions: A) phantom with the carbon cage inside it; B) phantom with the PET device switched off inside it; c) phantom with the PET device switched on inside it.

|   |   | Signal lossess in T3 (%) | Signal lossess in T10 (%) | Average signal losses (%) |
|---|---|---|---|---|
| (A-B)/A | slice 1 | 3.4 | -4.8 | |
|   | slice5 | 7.9 | -6.3 | 0.6 |
|   | slice9 | 9.9 | -6.6 | |
| (B-C)/B | slice 1 | 20.8 | 20.7 | |
|   | slice5 | 17.9 | 18.1 | 19.2 |
|   | slice9 | 18.8 | 20.1 | |
| (A-C)/A | slice 1 | 23.5 | 16.9 | |
|   | slice5 | 24.4 | 12.9 | 19.7 |
|   | slice9 | 26.8 | 13.7 | |

## 6. The endorectal transmitting and receiving RF coil

Endorectal MR imaging in local staging of prostate cancer is debated extensively in the literature [35]. Fütterer et al. [36] found that the area under a receiver operating characteristic curve would be significantly higher when endorectal coils were used compared with pelvic coils. Similarly, Turkbey et al. [37] found that more cancer foci were detected using dual-coil prostate MRI than when non-endorectal coil MRI was used at 3T. However, Bratan et al. [38] contests this claim with findings that the field strength and the type of imaging coils used have no significant influence on the detection rate of tumors.

MR surface RF coils can be adapted to the anatomy of interest and can be placed very close to the patient. One drawback of this type is the limited penetration depth and possible $B_1^+$-inhomogeneity. The penetration depth of a surface RF coil is approximately of the same order of magnitude as its diameter. Surface coils for an endorectal environment can be built either as receive only coils which only receive signal, or transmit and receive coils, which transmit and receive simultaneously. Considering the geometry and the limited space inside the endorectal probe, dedicated finite difference time domain calculations will be performed to describe the B1 fields in a uniform phantom and in a realistic model of the human prostate. The electrical and magnetic B1 fields based on different loop designs will be simulated. The use of multiple coil elements for transmission and receive of the MR signal has several benefits but reduces the sensitive area. Some studies have shown improved SNR and reduced RF power deposition at high field strength using by a stripline element to the endorectal RF loop transceiver for pelvic organs imaging [39].

The simplest- and most space-saving design is a single copper loop that is tuned and matched to the requirements of the MR system. Different approaches with varies order of the electronic components have been checked. Most commercial endorectal coils are receive-only. To get maximal output and optimal imaging quality a transmit-receive surface coil is a good option. A transceiver coil can improve SNR and decreased the prostate motion. A preliminary transmit and receive coil has been built by the Aachen University of Applied Sciences to demonstrate the performance. The electronic components on the coil are placed by smaller, also nonmagnetic, components. Further a capacitor on the backside copper area completes the shielding. The capacitor prevents the low frequency current flow on the shield of the cable caused by the gradient. A circuit board was planned and constructed to integrate the T/R-switch and low noise amplifier (see figure 10). In this case the coil consists of a FR4 plate with a single rectangular copper loop on top. The rectangular loop is 80 mm in length, 20mm in height and the wire diameter is 2mm. The first coil already shows low reflection and a good homogeneity.

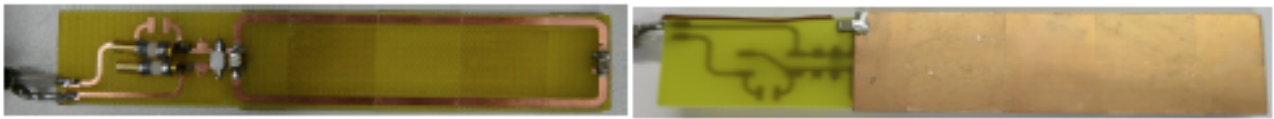

Figure 10 : a) Top side and b) backside of the preliminary prototype of the endorectal transmit and receive coil.

The system is adapted to different Siemens 3T MR systems with a frequency of 123,2 MHz and 50 Ω input. To further adapt the coil system to the different MR scanners the unique connector and cable have to be prepared. In our case systems for Siemens Allegra and Siemens Trio A Tim were built.

The preamplifier and the T/R-switch will be placed in a small case placed approximately 30cm away from the general set-up of the coil.

In the second iteration the distance between the copper shield on the bottom side of the FR4 plate and the surface loop on top was increased from 1.5mm up to 3.0mm (see figure 12). The shield is needed to reduce the interference of the PET electronics in the signal-to-noise ratio (SNR) of the MR image. The additional distance increases SNR by about 20%.

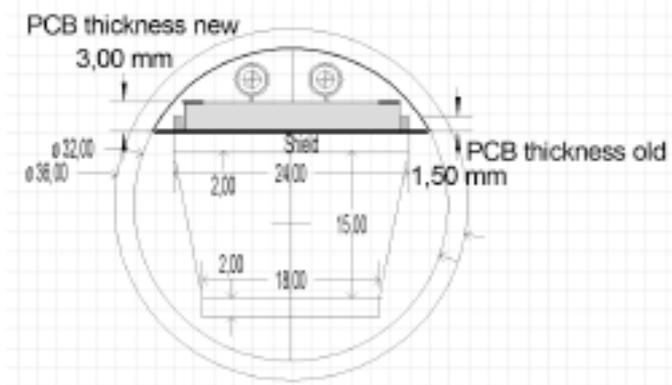

Figure 11: Transversal view of the integrated endorectal system: the FR4 plate with the copper loop on top and a shield on the bottom, PET detector module including SiPM and tapered crystal block in the centre of the tube.

For MR imaging the coil was placed in the phantom filled with foam material and placed on the table using a holder made of foam material. The orientation of the phantom was horizontal in the MR scanner (the Y-axis of the phantom matches the Z-axis of the MRI system).

The cylindrical ring at the transversal slice shows the PMMA phantom. The dark circle in the middle presents the probe. The coil is oriented upward, therefore the signal is stronger at the upper part of the shielding on the backside of the phantom. The corresponding images are taken by using MPRAGE sequences and show a significant improvement on the signal intensity. The corresponding SNR values increased from 35dB for the first coil design up to 42dB for the new design.

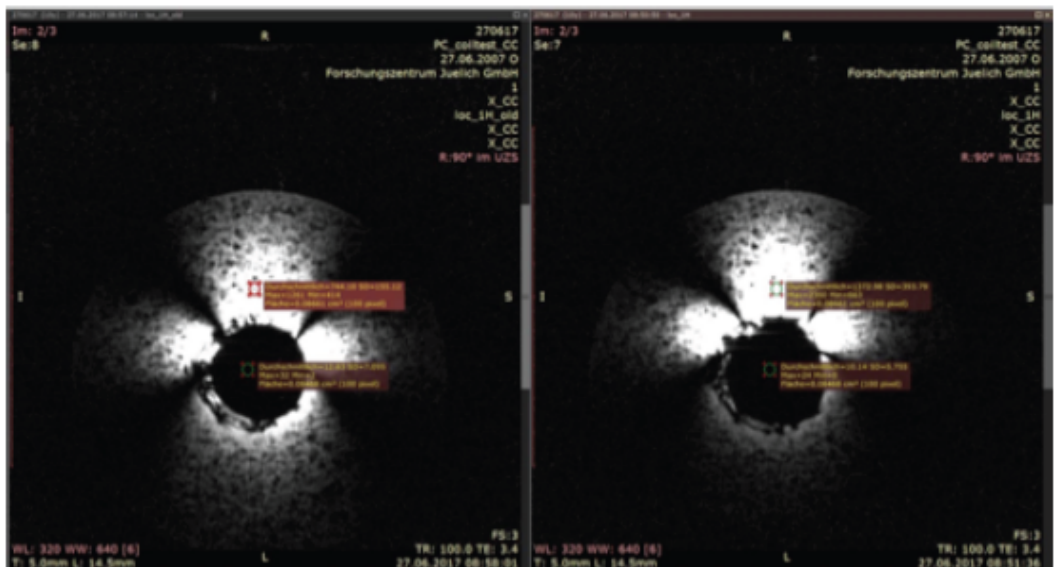

Figure 12: Transversal slices of a cylindrical phantom filled with foam material and the endorectal cylindrical tube including the transmit-receive coil at 3 T MRI-scanner, Siemens Trio A Tim at Research Center Jülich. The picture at left a) was done with the first version having 1.5mm distance between the shield and the loop, and at right b) with new design and larger distance of 3mm in between.

## 7. Summary, conclusions and future plans

We have developed a prototype of an endorectal PET-TOF probe, compatible with MRI, for diagnosis, biopsy, surgery guidance, and follow up of prostate cancer. The tests performed show that such a the detector is feasible and that the performance can be improved with fine tuning of different parameters and a careful choice of the materials to minimize the PET – MR adverse interference effects. Critical part of the project is the treatment of the scintillator pixel surfaces to obtain a good TOF resolution, very important parameter for the global performance of the diagnostic system, and at the same time good DOI. New collaborations have been established for the software development [40]. An high resolution ultrasound sensor [41] to be put into the probe will be needed to help guiding the biopsy procedure and also as third imaging modality for diagnosis purposes. Further progress in electronics is possible by designing a new ASIC [42], trying to reduce the intrinsic contribution of the front-end electronics to the timing errors down to a few ps rms, in order to have a resolution only limited by the sensor. Progress in SiPM technology, especially in low capacitance versions (for example 2 x 2 $mm_2$ pitch for the panels and 1 x 1 $mm^2$ for the probe), would allow further improvement in TOF [43], as an important parameter of the device. 100 ps for TOF resolution of such a system seems a reasonable target. Further designs of an MR coil are needed in order to improve the image quality and minimize interference between the PET-module and the MR imaging part.